\documentclass[aps,prb,twocolumn]{revtex4}
\usepackage{epsfig}
\usepackage{amsfonts}
\usepackage{latexsym}
\usepackage{epic}
\usepackage{graphics}
\usepackage{amsfonts}
\usepackage{graphicx}
\usepackage{epic,eepic,epsfig}
\newcommand{\re}[1]{(\ref{#1})}
\newcommand{\be}{\begin{equation}}
\newcommand{\ee}{\end{equation}}
\newcommand{\bea}{\begin{eqnarray}}
\newcommand{\eea}{\end{eqnarray}}
\newcommand{\beas}{\begin{eqnarray*}}
\newcommand{\eeas}{\end{eqnarray*}}
\newcommand{\no}{\nonumber}
\newcommand{\n}{{\vec{n}}}
\newcommand{\m}{{\vec{m}}}

\newcommand{\Th}{{\bf {\rm Th}}}

\newcommand{\ab}[1]{{\bf #1}}
\renewcommand{\S}{{\cal S}}
\newcommand{\T}{{\cal T}}

\begin{document}
\title{Hofstadter Problem on the Honeycomb and Triangular Lattices:
Bethe Ansatz Solution}
\author{M.~Kohmoto $^{a}\;$}
\email{kohmoto@issp.u-tokyo.ac.jp}

\author{A.~Sedrakyan$^{a,b}\;$}
\email{sedrak@alf.nbi.dk}

\affiliation{$^a$ Institute of Solid State Physics, University of Tokyo,\\
Kashiwanoha 5-1-5, Kashiwa-shi, Chiba, 277-8581, Japan\\
$^b$ Yerevan Physics Institute,Br.Alikhanian str.2,
Yerevan 36, Armenia}
\date{\today}

\begin{abstract}
We consider Bloch electrons on the honeycomb lattice under a uniform
magnetic field with $2 \pi p/q$ flux per cell.
It is shown that the problem factorizes to two  triangular
lattices. Treating magnetic translations as Heisenberg-Weyl group and by the use
of its irreducible representation on the space of theta functions, we find a nested
set of Bethe equations, which determine the eigenstates and energy spectrum.
The Bethe equations have simple form which allows to consider them further
in the limit $p, q \rightarrow \infty$ by the technique of Thermodynamic Bethe Ansatz
and analyze Hofstadter problem for the irrational flux.
\end{abstract}
\maketitle

\newpage
\section{Introduction}
\indent
The quantum Hall effect (QHE) on graphite thin films has been observed recently \cite{Nov, Zheng} .
It has the honeycomb lattice with one $\pi$ electron per lattice site. The only Fermi
levels ($E_F=0$) are two points of the Brillouin zone where the conduction
band and the valence bands touch forming cones. Low energy excitations have a
linear spectrum as for the massless relativistic Dirac particles in 2+1 dimensional
space-time. \cite{J, Sem, Haldain}. In \cite{Ando, Gusinin, Peres}
the  Hall effect on graphene on the basis of Dirac electrons was investigated.

The problem of electrons in two dimensional periodic potential in a
magnetic field has been attracted a lot of attention
due to the unexpected nature of the commensurability and frustrations.
\cite{Azbel,Zak, Wannier, Hofstadter, TKNN,KAP,K1,Hiramoto,K2,K3,K4,Wiegmann,HS}.
It has let the existence of Cantor set energy spectra, and the wavefunctions strongly
depend on wether the flux is commensurate or incommensurate with the lattice.
It was observed \cite{Hofstadter} that when the magnetic flux corresponding to
unit cell of the periodic structure of the system is incommensurate with the potential,
the one particle spectrum exhibits multifractal behavior like the Cantor set
(see also \cite{Hiramoto}). Then, after discovering the topological character of the
conductance in the model by linking it with the Chern class in the linear bundle of the
Bloch wave function \cite{TKNN,KAP}, its possible connection with quantum Hall effect
become evident.

Usually a solid with well localized atomic orbits is modelled by the
lattice and the investigations were carried out by the use of a
tight-binding Hamiltonian..
The presence of zero modes was pointed out in \cite{K1} and the
model with nearest-neighbor and next-to-nearest-neighbor hopping was studied in
\cite{K2}. Wiegmann and Zabrodin \cite{Wiegmann} have observed the presence of
a hidden quantum group in the Hofstadter problem on the regular lattice and
initiated studies \cite{K4, Faddeev, HS} of the integrable structure in it. The
approach allowed to express the spectrum and the Bloch wave function at the
mid-band points as solutions of the Bethe equations typical for completely
integrable quantum systems. In the article \cite{Faddeev} the problem have been
analyzed by the use of cyclic representations of quantum groups at any momenta, but the
explicit form of the Bethe equations was not found(besides the midband point).

In \cite{HS} authors have used the fact that magnetic translations
in Hofstadter problem are forming Heisenberg-Weyl group and analyzed its
irreducible representations on the basis of theta functions \cite{Manford}.
This approach allowed us to find explicit form of the Bethe equations for the spectrum
and the wave-functions not only at the midband point, but for any momenta.

Motivated by the discovered unconventional Hall effect in graphene we study
in this article the  problem of Bloch electrons in a uniform magnetic field with
flux $2 \pi p/q$ per cell on the honeycomb lattice. We show that the problem
possesses a chiral factorization and it reduces to two ($R$ and $L$) triangular
sublattices. Following the
line of \cite{HS}, we look for the solution of the
Schr\"odinger equation for energy and eigenfunctions in the space of
 irreducible representation of the Heisenberg-Weyl group, namely, in the
space of theta functions with characteristics.
By the use of nesting procedure, similar to the one appeared in Algebraic Bethe
Ansatz for some of integrable models (see for example \cite{EK, EK1, TS}), we
have transformed Schr\"odinger equation into the
simple set of nested Bethe equations for the spectral parameters.
We have found a spectrum of the model and eigenstates as a function
of the solution of the nested Bethe Anzats equations, which depend on momenta
$\vec k$.

The form of the nested Bethe equations allow to investigate them in thermodynamic
limit $q \rightarrow \infty$ by the technique developed by Takahashi
\cite{Takahashi} and Gaudin \cite{Gauden}  for integrable models and called
Thermodynamic Bethe Ansatz. This limit is important for the most interesting
irrational flux case of the Hofstadter problem since $q$ gives the number of bands
and we face a Cantor like behavior of the spectrum.
Irrational number can be approached by the rationales $p/q$ with
$p, q \rightarrow \infty$.

\section{Electrons on the 2d honeycomb lattice in a uniform magnetic filed}
\indent

\begin{figure}[ht]
\centerline{\includegraphics[width=90mm,angle=0,clip]{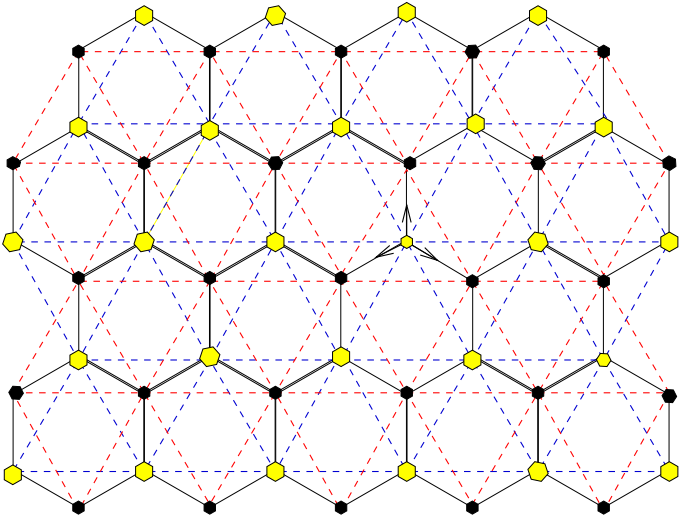}}
\caption{The representation of the honeycomb lattice as a joint
of triangular ${\cal L}^L$ (red) and ${\cal L}^R$ (blue) lattices.}
\label{fig:1}
\end{figure}

Let  us start with formulation of Hofstadter problem on the
honeycomb lattice $\cal L$ (see FIG.1). The honeycomb lattice possesses chiral
factorization, namely one can consider $\cal L = {\cal L}^L+ {\cal L}^R $ as a
sum of two triangular lattices ${\cal L}^L$ and ${\cal L}^R$, each of which is
isomorphic to the dual of the honeycomb lattice. Sites of the lattices ${\cal
L}^L$ and ${\cal L}^R$ on the Fig.\ref{fig:1} are marked with black and yellow
dots respectively. The links of the honeycomb lattice are connecting the sites
on ${\cal L}^R$ with the three nearest neighbor sites on  ${\cal L}^R$.

The Hamiltonian is given by
\bea
\label{Hamil}
H &=&\sum_{\n \in {\cal L}^R, \sigma={1,2,3}} \big(t_{\n,\n+{\vec e}_{\sigma}}
e^{iA_{\n,\n+{\vec e}_{\sigma}}}c_\n^+d_{\n+{\vec e}_{\sigma}} \no \\
&+&t_{\n+{\vec e}_{\sigma},\n}
e^{iA_{\n+{\vec e}_{\sigma},\n}}d_{\n+{\vec e}_{\sigma}}^+c_\n\big),
\eea
where $c_\n,c_\n^+$ and $d_{\n+{\vec e}_{\sigma}},d_{\n+{\vec e}_{\sigma}}^+$ are
the annihilation and creation operators of electrons
at the sites $\n$ and $\n+{\vec e}_{\sigma}$ of the triangular lattices
${\cal L}^R$ and ${\cal L}^L$ respectively.
 $A_{\n,\n+{\vec e}_{\sigma}}=-A_{\n+{\vec e}_{\sigma},\n}$ is the vector potential
of a  magnetic  field with the  strength perpendicular to the plane of the
lattice and $t_{\n,\m}$  are hoping   amplitudes between the nearest neighbors. We
choose  them as $t_1$, $t_2$ and  $t_3$ according to directions of the vectors
${\vec e}_{\sigma}$ (see Fig.2). For the homogeneous model $t_1=t_2=t_3$.

In the following we consider
the diagonalization problem of \re{Hamil} in one-particle sector.
In this sector the action of the translation operator by vectors $\pm \vec{e}_{\sigma}$
can be written as
\bea
\label{tra}
S_{\vec{e}_{\sigma}}=\sum_{\n \in {\cal L}^R }\vert\n\rangle\langle\n+\vec{e}_{\sigma}\vert,\quad
S_{-\vec{e}_{\sigma}}=\sum_{\n \in {\cal L}^R }\vert\n+\vec{e}_{\sigma}\rangle\langle\n\vert,
\eea
where the standard bra- and ket- vectors are used:
$\vert\n \rangle=c_\n^+|vac\rangle$,
 $\langle\n\vert=\langle vac |c_\n$ and
$\vert\n+\vec{e}_{\sigma} \rangle=d_{\n+\vec{e}_{\sigma}}^+|vac\rangle$,
$\langle{\n+\vec{e}_{\sigma}}\vert=\langle vac |d_{\n+\vec{e}_{\sigma}}$.

We consider  the case when the magnetic  flux per hexagon cell,
\begin{equation}
\label{plaq}
\exp(i\Phi)=\prod_{hexagon} \exp(iA_{\n,\m}),
\end{equation}
 is rational:
$\Phi=2\pi {p\over q}$, where $p$ and $q$ are mutually prime  integers.
 The product in  \re{plaq} is performed on anticlockwise
direction.

In terms of bra- and ket- vectors the Hamiltonian (\ref{Hamil}) can be written as
\bea
\label{Hamil-2}
H&=&\sum_{\n \in {\cal L}^R,\atop \sigma={1,2,3}} \big(t_{\n,\n+{\vec e}_{\sigma}}
e^{iA_{\n,\n+{\vec e}_{\sigma}}}\vert \n \rangle \langle \n+{\vec e}_{\sigma} \vert \no\\
&+& t_{\n+{\vec e}_{\sigma},\n} e^{iA_{\n+{\vec e}_{\sigma},\n}}\vert \n+{\vec e}_{\sigma}\rangle
\langle \n \vert\big)+ h.c.
\eea
\begin{figure}[t]
\centerline{\includegraphics[width=50mm,angle=0,clip]{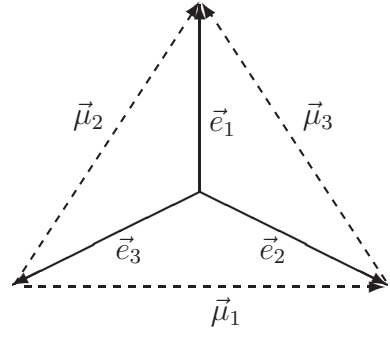}}
\caption{Hopping vectors on the honeycomb ($\vec{e}_{\sigma}$) and triangular
($\vec{\mu}_{\sigma}$) lattices}
\label{vec}
\end{figure}

Let us now consider the square of the Hamiltonian (\ref{Hamil-2}). It is
straightforward to see from the expression (\ref{Hamil-2}) that $H^2$
contain only hopping terms within
triangular lattices ${\cal L}^R$ and ${\cal L}^L$, namely
\begin{eqnarray}
\label{Hamil3}
H^2 = {\cal H}_R+ {\cal H}_L,
\end{eqnarray}
where
\begin{eqnarray}
\label{HRL}
 {\cal H}_{R(L)}&=& \sum_{\n \in {\cal L}^R({\cal L}^L),\atop \sigma =1,2,3}
({\tilde t}^{R(L)}_{\n,\n+{\vec \mu}_{\sigma}}
e^{iA_{\n,\n+{\vec \mu}_{\sigma}}}\vert \n \rangle \langle \n+{\vec \mu}_{\sigma} \vert
+h.c.)\no\\
&+&\sum_{\n \in {\cal L}^R({\cal L}^L)}
{\tilde t}_{\n,\n}\vert \n \rangle \langle \n \vert,
\end{eqnarray}
with (see FIG. 2)
\begin{eqnarray}
\label{par}
{\vec \mu}_{1}={\vec e}_2 -{\vec e}_3,&&
{\vec \mu}_{2}={\vec e}_1 -{\vec e}_3, \no\\
{\vec \mu}_{3}&=&{\vec e}_1 -{\vec e}_2,
\end{eqnarray}
and
\begin{eqnarray}
\label{t}
{\tilde t}^{R}_{\n,\n\pm{\vec \mu}_{1}}=e^{\mp i \frac{\Phi}{6}}t_2 t_3,&&
{\tilde t}^{R}_{\n,\n\pm{\vec \mu}_{2}}=e^{\pm i \frac{\Phi}{6}}t_1 t_3,\no\\
{\tilde t}^{R}_{\n,\n\pm{\vec \mu}_{3}}=e^{\mp i \frac{\Phi}{6}}t_1 t_2,&&
{\tilde t}_{\n,\n}=t^2_1+ t^2_2+t^2_3\no\\
 {\tilde t}^{L}&=&\big({\tilde t}^{R}\big)^*
\end{eqnarray}


This decomposition reveal the hidden chiral structure of the Hofstadter problem on
the honeycomb lattice reducing it to the problem on the triangular lattice.
Therefore, in the remaining part of this article, we will consider only the triangular
lattice.

The problem of Bloch electrons on the triangular lattice and in uniform magnetic
filed was considered recently in \cite{Ino}.
\vspace{0.5 cm}
\section{Electrons on the triangular lattice and in uniform magnetic field}
\indent

Denote lattice sites by vectors $\n = n_1 {\vec \mu}_1 + n_2 {\vec \mu}_2$.
Taking different  gauges one  can obtain various equivalent forms of
\re{Hamil3}. In the following we will use the Landau gauge, which  is
\begin{eqnarray}
\label{Land}
&&A_{\pm {\vec \mu}_1}(\n)=A_{\n,\n \pm \vec{\mu}_1}=0, \no \\
&&A_{\pm {\vec \mu}_2}(\n)=A_{\n,\n \pm \vec{\mu}_2}=\pm \Phi n_1.
\end{eqnarray}
Then the Hamiltonian (\ref{Hamil3}) is invariant under translations
$ S_{\pm {\vec \mu}_1}^q$,\ $S_{\pm {\vec \mu}_2}$ and $S_{\pm {\vec \mu}_3}^q$:
$$
{[}{\cal H},S_{\pm {\vec \mu}_1}^q{]}={[}{\cal H},S_{\pm {\vec \mu}_2}{]}=
{[}{\cal H},S_{\pm {\vec \mu}_3}^q{]}=0.
$$
So, the problem  of diagonalization of ${\cal H}$ reduced to its
diagonalization on each  eigenspace of $S_{\pm {\vec \mu}_1}^q$, $S_{\pm {\vec
\mu}_2}$ and $S_{\pm {\vec \mu}_3}^q$. The latter is a $q$-dimensional  space
$\Psi(\vec{k})$, spanned by Bloch wave functions
$$
\psi_{n}(\vec{k}) = \sum_{\n_1=n {\rm mod}q \atop \n_2} e^{-i \vec{k}\n}
 \vert\n\rangle,
$$
which   satisfy
\bea
\label{trrep}
& \psi_{n}(\vec{k}) =\psi_{n+q}(\vec{k}) &
\no\\
& S_{\pm {\vec \mu}_1}\psi_{n}(\vec{k})=e^{\mp ik_1}\psi_{n\mp 1}(\vec{k}) &\\
& S_{\pm {\vec \mu}_2}\psi_{n}(\vec{k})=e^{\mp ik_2}\psi_{n}(\vec{k}) & \no \\
& S_{\pm {\vec \mu}_3}\psi_{n}(\vec{k})=e^{\mp i(k_2-k_1)}\psi_{n\pm 1}(\vec{k}). &\no
\eea
Let us define also the generators  of magnetic translations by
\begin{eqnarray}
\label{mag}
T_{\pm \vec{\mu}_{\sigma}}=\sum_\n e^{iA_{\n,\n\pm \vec{\mu}_\sigma}}\vert\n\rangle
  \langle\n\pm \vec{\mu}_\sigma \vert, \quad \sigma=1,2,3.
\end{eqnarray}
It is easy to find out, that in Landau gauge (\ref{Land})
\begin{eqnarray}
\label{mag2}
T_{\pm \vec{\mu}_{3}}=e^{i \Phi/2}T_{\mp \vec{\mu}_{1}} T_{\pm \vec{\mu}_2},
\end{eqnarray}
since the area of the triangles in honeycomb lattice (Fig.1) is the half of the
area of the hexagon.

The generators $T_{\pm \vec{\mu}_{1}}$ and $T_{\pm \vec{\mu}_{2}}$ satisfy the
following  commutation relations
\bea
\label{Heisenberg}
T_{\pm \vec{\mu}_{1}}T_{\pm \vec{\mu}_{2}}=\bar q^{2}T_{\pm \vec{\mu}_{2}}T_{\pm \vec{\mu}_{1}},
\eea
and  form   Heisenberg-Weyl group. Here we used the notation
$$
\bar q=\exp\left(i\frac{\Phi}{2}\right)=\exp\left(\pi i\frac{p}{q}\right).
$$
Note, that  is any gauge $S_{\vec{\mu}_{1},\vec{\mu}_{2}}^q=T_{\vec{\mu}_{1},\vec{\mu}_{2}}^q$
(in the Landau gauge moreover $S_{\vec{\mu}_{1}}=T_{\vec{\mu}_{1}}$).

The action of magnetic translations $T_{\pm \vec{\mu}_{1}}$, $T_{\pm \vec{\mu}_{2}}$
  on Bloch functions $\psi_{n}(\vec{k})$ has the following form:
\bea
\label{magrep}
& T_{\pm \vec{\mu}_{1}}\psi_{n}(\vec{k})=e^{\mp ik_1}\psi_{n\mp 1}(\vec{k}) & \no \\
&T_{\pm \vec{\mu}_{2}}\psi_{n}(\vec{k})=e^{\mp ik_2\pm i n \Phi}\psi_{n}(\vec{k}) &  \\
&T_{\pm \vec{\mu}_{3}}\psi_{n}(\vec{k})=e^{\mp i(k_2-k_1)\pm i (n\pm 1/2) \Phi}\psi_{n\pm 1}(\vec{k}). & \no
\eea
In case of translational invariant distribution of hopping parameters
$\tilde{t}_{\pm \sigma}={\tilde t}_{\n,\n\pm{\vec \mu}_{\sigma}},\;\sigma=1,2,3$,
and $\tilde{t}={\tilde t}_{\n,\n}$,
the Hamiltonian \re{Hamil} can be written in terms of  the generators of
magnetic translations by
\begin{eqnarray}
\label{hamil4}
{\cal H}=\sum_{\sigma=1,2,3}({\tilde t}_{\sigma}T_{\vec{\mu}_{\sigma}}+
{\tilde t}_{-\sigma}T_{-{\vec{\mu}_{\sigma}}})+\sum_{\n}
{\tilde t}\vert \n \rangle \langle \n \vert.
\end{eqnarray}
Note that on $\Psi(\vec{k})$  the $q$-th power  of magnetic translations
are scalars:
\bea
\label{center}
 T_{\pm {\vec{\mu}_{1}}}^q =e^{\mp iqk_1}\cdot {\rm id}, && T_{\pm {\vec{\mu}_{2}}}^q = e^{\mp iqk_2}
\cdot {\rm id}, \no\\
T_{\pm {\vec{\mu}_{3}}}^q &=& e^{\mp iq(k_1-k_2)}\cdot {\rm id}
\eea
In order to solve the Schr\"odinger of equation ${\cal H}\psi={\cal E}\psi$
analiticaly we will follow the
technique developed in \cite{HS} for the Hofstadter problem on the regular lattice.

\section{Heisenberg-Weyl group. Reduction of the problem to the Bethe equations}
\indent

Let us recall
the  representation of Heisenberg-Weyl group \re{Heisenberg} on  the space of
complex functions \cite{Manford}.
It can be constructed in the following way.
Define actions $\S_b$ and $\T_a$, $a,b\in Z$
on the space of analytic functions on the complex  plain  as
\bea
\S_bf(z)&=&f(z+b), \\
\T_a(\tau)f(z)&=&\exp(\pi ia^2\tau+2\pi iaz) f(z+a\tau),
\no
\eea
for some $\tau\in C$. Then
\bea
\S_a\S_b &=&\S_{a+b},  \T_a(\tau)\T_b(\tau)=\T_{a+b}(\tau)\no\\
\S_a\T_b(\tau) &=& \exp(2\pi iab)\T_b(\tau)\S_a.
\eea
Consider the space of theta functions with characteristics
$\Theta^{(q)}(z,\tau)$
which are invariant with respect to the subgroup generated by
$\S_{\pm q}$ and $\T_{\pm 1}(\tau)$. They form a q-dimensional space
$\Th_q(\tau)$
and have in the fundamental domain with vertices $(0,\tau,q,q+\tau)$
precisely $q$ zeroes.

The space $\Th_q(\tau)$
forms an irreducible representation of Heisenberg-Weyl group \re{Heisenberg}
generated by
\bea
\label{T_x}
 T_{\pm {\vec \mu}_1}&=& \bar\alpha_\pm \T_{\mp {1\over q}}(\tau), \quad
 T_{\pm {\vec \mu}_2}=\bar\beta_\pm \S_{\pm{p}}=\bar\beta_\pm \S_{\pm{1}}^p, \no\\
T_{\pm {\vec \mu}_3} &=& \bar\alpha_\mp \bar\beta_\pm e^{i \Phi/2} \T_{\pm {1\over q}}(\tau) \S_{\pm{p}} .
\eea
One can choose the basis $\Theta_{a,0}(z,\tau)$ of $\Th_q$ as follows:
$$
\Theta_{a,0}(z,\tau)=\Big[\T_{a}(\tau)\Theta\Big](z,\tau),
$$
where $ a= 0,1/q,\dots,(q-1)/q$,
$\Theta(z,\tau)=\Theta_0(z,\tau)$ is the standard theta function
and $\Theta_{a,b}(z,\tau)=T_aS_b\Theta(z,\tau), \; \; a\in {1\over N}\ab{Z} \
{\rm mod 1}, \; b\in {1\over M}\ab{Z} \ {\rm mod 1}$
(\cite{Manford}) is the notation for theta functions with
characteristics $\left[\begin{array}{c}
N\\
M
\end{array}\right]$.
In this basis
\beas
\S_{\pm 1}\Theta_{a,0}=\exp(\pm {2\pi ia})\Theta_{a,0}
\no\\
\T_{\pm{1\over q}}\Theta_{a,0}=\Theta_{a\pm {1\over q},0}
\eeas
Comparing these equations with \re{magrep} and \re{center}
we obtain for parameters $\bar\alpha_\pm,\, \bar\beta_\pm$
\bea
&\bar\alpha_\pm=\exp(\mp ik_1) & \bar\beta_\pm =\exp(\mp ik_2), \no
\eea
The space $\Psi(\vec{k})$ is identified with the space of theta
functions with characteristic $\left[\begin{array}{c}
q\\
1
\end{array}\right]$:
$$
\psi_{n}(\vec{k})\sim \Theta_{{n\over q},0}(z,\tau)
$$

The space $\Th_q(\tau)$ have some analog with the space of polynomials
of degree $q-1$. The polynomial decomposition analog for the $\Theta^{(q)}$
is the following.
Every $\Theta^{(q)} \in \Th_q$ can be represented in the
form
\bea
\label{basis}
\Theta^{(q)}(z,\tau)=\gamma\exp\left(-\frac{2\pi ikz}{q}\right)
\prod_{j=1}^q\Theta\left(\frac{z-z_j}q,\frac\tau q\right)
\eea where $ \sum_{i=1}^q z_i=k\tau+nq, \ \ k,n\in\ab{Z}, \ \ 0\le k,n <q, \ \
\\ \gamma\in \ab{C}$.
Then for our Hamiltonian on triangular lattice
\begin{eqnarray}
\label{calH}
{\cal H}&=&  t_2t_3 \big(e^{-i\frac{\Phi}{6}}T_{{\vec \mu}_1}+
 e^{i\frac{\Phi}{6}}T_{{\vec \mu}_{-1}}\big)
+t_1 t_3\big(e^{i\frac{\Phi}{6}}T_{{\vec \mu}_2}+\no\\
&+& e^{-i\frac{\Phi}{6}}T_{-{{\vec \mu}_2}}\big) +t_1 t_2 \big(e^{-i\frac{\Phi}{6}}T_{{\vec
\mu}_3} + e^{i\frac{\Phi}{6}}T_{-{{\vec \mu}_3}}\big),
\end{eqnarray}
which was induced from the Hofstadter problem on the honeycomb lattice and the
Schr\"odinger equation
 ${\cal H}\psi={\cal E}\psi$ can be written in terms
 of theta functions. It is convenient to write them in terms of
$$
\Theta_1(z,\tau)=\exp\left({\pi i\tau \over 4}+{\pi i\over 2}- \pi i z\right)
\Theta\left(z+{1-\tau \over 2},\tau\right),
$$
which obey $\Theta_1(0,\tau)=0$. After rescaling of the arguments of
$\Theta_1(z,\tau)$: $z\to qz$, $\tau\to q\tau$ and performing additional
substitution: $z\to z+{{1+\tau}\over 2}$ we obtain
\begin{widetext}
\bea \label{eq:B}
 &&(-1)^p\beta_+ e^{i\frac{\Phi}{6}-\frac{2\pi i k p}{q}}
    \prod_{i=1}^q \Theta_1\Big(z-z_i+{p\over  q},\tau\Big)
 + (-1)^p\beta_-e^{-i\frac{\Phi}{6}+\frac{2\pi i k p}{q}}\prod_{i=1}^q
    \Theta_1\Big(z-z_i-{p\over  q},\tau\Big) \no\\
&+&\alpha_+ e^{-i\frac{\Phi}{6}+\pi i\tau\frac{-q+1+2k}q-2\pi i z}
        \prod_{i=1}^q \Theta_1\Big(z-z_i-{\tau\over  q},\tau\Big)
+\alpha_-e^{i\frac{\Phi}{6}+ \pi i\tau\frac{q+1-2k}q+2\pi i z}
        \prod_{i=1}^q \Theta_1\Big(z-z_i+{\tau\over  q},\tau\Big)
\no\\
&+&(-1)^p\delta_+e^{i\frac{\Phi}{3}+\pi i\tau\frac{1-2k+q}{q}+2\pi i z
-\frac{2\pi i k p}{q}}
        \prod_{i=1}^q \Theta_1\Big(z-z_i+{\tau+p \over  q},\tau\Big) \\
&+& (-1)^p\delta_-e^{i\frac{2\Phi}{3}+\pi i\tau\frac{1+2k-q}{q}-2\pi i z
+\frac{2\pi i k p}{q}}
        \prod_{i=1}^q \Theta_1\Big(z-z_i-{\tau+p \over  q},\tau\Big)
= ({\cal E}-3)\prod_{i=1}^q \Theta_1\Big(z-z_i,\tau\Big) \no
\eea
where
\begin{eqnarray}
\label{newaplha}
\alpha_{\pm}=t_2t_3\bar\alpha_{\pm}=t_2t_3e^{\mp i k_1},\;
\beta_{\pm}=t_1t_3\bar\beta_{\pm}=t_1t_3e^{\mp i k_2},\;
\delta_{\pm}=t_1t_2\bar\alpha_{\mp}\bar\beta_{\pm}=t_1t_2e^{\pm i (k_1-k_2)}
\end{eqnarray}
\end{widetext}
and ${\cal E}$ is the eigenvalue of the Hamiltonian
(\ref{calH}) on the triangular lattice.

The roots of right and left side of (\ref{eq:B}) must
coincide. So, inserting the zeros $z=z_i, \  i=1,\dots,q$ of $\Theta_1$
into the left hand side we
obtain the system of $q$ equations for the spectral parameters $z_i$:
\begin{widetext}
\bea \label{BAE1}
 &&(-1)^p\beta_+ e^{i\frac{\Phi}{6}-\frac{2\pi i k p}{q}}
    \prod_{i=1}^q \Theta_1\Big(z_j-z_i+{p\over  q},\tau\Big)
 + (-1)^p\beta_-e^{-i\frac{\Phi}{6}+\frac{2\pi i k p}{q}}\prod_{i=1}^q
    \Theta_1\Big(z_j-z_i-{p\over  q},\tau\Big)\no\\
&+&\alpha_+ e^{-i\frac{\Phi}{6}+\pi i\tau\frac{-q+1+2k}q-2\pi i z_j}
        \prod_{i=1}^q \Theta_1\Big(z_j-z_i-{\tau\over  q},\tau\Big)
+\alpha_-e^{i\frac{\Phi}{6}+ \pi i\tau\frac{q+1-2k}q+2\pi i z_j}
        \prod_{i=1}^q \Theta_1\Big(z_j-z_i+{\tau\over  q},\tau\Big)
\no\\
&+&(-1)^p\delta_+e^{i\frac{\Phi}{3}+\pi i\tau\frac{1-2k+q}{q}+2\pi i z_j
-\frac{2\pi i k p}{q}}
        \prod_{i=1}^q \Theta_1\Big(z_j-z_i+{\tau+p \over  q},\tau\Big) \\
&+& (-1)^p\delta_-e^{i\frac{2\Phi}{3}+\pi i\tau\frac{1+2k-q}{q}-2\pi i z_j
+\frac{2\pi i k p}{q}}
  \prod_{i=1}^q \Theta_1\Big(z_j-z_i-{\tau+p \over  q},\tau\Big)
=0 \no
\eea
\end{widetext}

This is an analog of Bethe Ansatz equation for the problem under
consideration, but it has no convinient form for further investigations
and needs to be simplified.

By the use of the nesting procedure
in the Appendix A we show that the Schr\"odinger equation
(\ref{eq:B}) transforms to the very simple form
\begin{eqnarray}
\label{eq:Bel}
({\cal E}-3)
&=&\bar\gamma \frac{\prod_{i=1}^q \Theta_1\big(z-z^{(12)}_i,\tau\big)}
{\prod_{i=1}^q \Theta_1\big(z-z_i,\tau\big)}\no\\
&+&\gamma_{3} \frac{\prod_{i=1}^q \Theta_1\big(z-z^{(3)}_i,\tau\big)}
{\prod_{i=1}^q \Theta_1\big(z-z_i,\tau\big)},
\end{eqnarray}
where parameters $\gamma_3$ and $\bar\gamma$ are defined by the
formulas (\ref{gamafinal} - \ref{12}),
while the set of Bethe equations (\ref{BAE1}) is
equivalent to the following larger, but much simpler set of
nested Bethe equations
\begin{eqnarray}
\label{beta1} \frac{
    \prod_{i=1}^q \Theta_1(z^{(1)}_j-z_i+{p\over  q},\tau)}
    {\prod_{i=1}^q
    \Theta_1(z^{(1)}_j-z_i-{p\over  q},\tau)}
 =-\frac{\beta_-}{\beta_+ }e^{-i\frac{\Phi}{3}+\frac{4\pi i k p}{q}},\qquad
 \quad
\end{eqnarray}
\begin{eqnarray}
\label{alpha1} \frac{
        \prod_{i=1}^q \Theta_1(z^{(2)}_j-z_i-{\tau\over  q},\tau)}
        {\prod_{i=1}^q \Theta_1(z^{(2)}_j-z_i+{\tau\over  q},\tau)}=
\qquad \qquad \qquad \qquad \qquad \no\\
=-\frac{\alpha_-}{\alpha_+} e^{i\frac{\Phi}{3}+ 2\pi i \tau\frac{q-2k}{q}+4\pi i
z^{(2)}_j},\qquad
\end{eqnarray}
\begin{eqnarray}
\label{alphabeta1} \frac{
 \prod_{i=1}^q \Theta_1(z^{(3)}_j-z_i+{\tau+p\over q},\tau)}
 {\prod_{i=1}^q \Theta_1(z^{(3)}_j-z_i-{\tau+p \over q},\tau)}=
\qquad \qquad \qquad \qquad \quad \no\\
=-\frac{\delta_-}{\delta_+}e^{i\Phi(\frac{1}{3}+2 k)-
2\pi i \tau\frac{q-2 k}{q}-4\pi i z^{(3)}_j},\qquad
\end{eqnarray}
\begin{eqnarray}
\label{gamagama}
e^{i\frac{\Phi}{3}-\pi i\tau\frac{3-q+2k}{q}
-\frac{2\pi i k p}{q}+2\pi i z_1} \frac{\prod_{i=1}^q \Theta_1(z^{(12)}_j-z^{(1)}_i,\tau)}
{\prod_{i=1}^q \Theta_1(z^{(12)}_j-z^{(2)}_i,\tau)} \quad \no\\
=(-1)^{p+1}\frac{\alpha_+}{\beta_+}\frac{\prod_{i=1}^q \Theta_1(z_1-z_i-{2\tau\over q},\tau)}
 {\prod_{i=1}^q \Theta_1(z_1-z^{(2)}_i-{\tau\over q},\tau)}
\quad \no\\
\cdot
\frac{\prod_{i=1}^q \Theta_1(z_1-z^{(1)}_i+\frac{p}{q},\tau)}
{\prod_{i=1}^q \Theta_1(z_1-z_i+\frac{2p}{q},\tau)}\quad
\end{eqnarray}

and
\begin{eqnarray}
\label{final1}
e^{-i\frac{7 \Phi}{6}-\pi i\tau\frac{3-2k+q}{q}-2\pi i z_1
} \frac{\prod_{i=1}^q \Theta_1(z_j-z^{(12)}_i,\tau)}
{\prod_{i=1}^q \Theta_1(z_j-z^{(3)}_i,\tau)}\no\\
= -\frac{\delta_+}{\beta_+}\frac{\prod_{i=1}^q \Theta_1(z_1-z_i+{2(p+\tau)\over q},\tau)}
 {\prod_{i=1}^q \Theta_1(z_1-z^{(3)}_i+{(p+\tau)\over q},\tau)}
\quad \no\\
\cdot
\frac{\prod_{i=1}^q \Theta_1(z^{(2)}_1-z^{(12)}_i,\tau)
\cdot \Theta_1(z_1-z^{(1)}_i+{p\over q},\tau)}
{\prod_{i=1}^q \Theta_1(z^{(2)}_1-z^{(1)}_i,\tau)
\cdot \Theta_1(z_1-z_i+{2 p\over q},\tau)}.\;\;
\end{eqnarray}
We have obtained a set of five equations for the set of five parameters
$z_i, z^{(1)}_i, z^{(2)}_i,  z^{(3)}_i, z^{(12)}_i ,\; i= 1,2,\cdots q$.
The parameters $\alpha_{\pm}, \; \beta_{\pm}$ and
$\delta_{\pm}$ depend on momenta $\vec k$. Therefore the solutions
$z_i, z^{(1)}_i, z^{(2)}_i,  z^{(3)}_i, z^{(12)}_i ,\; i= 1,2,\cdots q$
of the equations (\ref{beta1})-(\ref{final1}) depend on the
momenta too.

Both terms in the sum on the left hand side of (\ref{eq:Bel}) are elliptic
functions, i.e. 2-periodic, in the fundamental domain with vertexes
$(0,\tau,1,1+\tau)$, and each of them has precisely $q$ zeroes and $q$ poles
there. Let us suggest that all poles are single. It is known that such an
elliptic function can be written in the form

\be
f(z)=C+\sum_{i=1}^{q}A_i\zeta(z-z_i),
\label{eq:decom}
\ee
where $\zeta(z)$ is a $\zeta$-function of Weierstrass for the same periodicity
domain. Formula (\ref{eq:decom}) is an analog of decomposition of rational
functions with single poles in terms of functions $\frac{1}{z-z_i}$.

Applying (\ref{eq:decom}) like decomposition for elliptic functions to
the expression (\ref{eq:Bel}) for the energy one can verify that the conditions
for the annulation of the residues of the poles
precisely are coinciding with (\ref{beta1})-(\ref{final1})
(or equivalently (\ref{BAE1})).
Therefore the energy $E$ is independent of $z$ constant function $C$ in
this decomposition and one can find (see formulas (\ref{final}) and
(\ref{gama3}) in the Appendix)
\begin{eqnarray}
\label{final0}
{\cal E} =3+
(-1)^p\delta_+e^{i\Phi(\frac{4}{3}-k)+\pi
i\tau\frac{3-2k+q}{q}+2\pi i z_1} \qquad \\
 \cdot\frac{\prod_{i=1}^q \Theta_1(z_1-z_i+{2(p+\tau)\over q},\tau)
\cdot \Theta_1(z^{(12)}_1-z^{(3)}_i,\tau)}
 {\prod_{i=1}^q \Theta_1(z_1-z^{(3)}_i+{(p+\tau)\over q},\tau)
\cdot \Theta_1(z^{(12)}_1-z_i,\tau)},\no
\end{eqnarray}
This expression defines the spectrum of the states in the
Hofstadter problem on the honeycomb lattice as a function of momenta
because the solutions $z_i,\; z^{(1)}_i,\;
z^{(2)}_i,\;  z^{(3)}_i,\; z^{(12)}_i ,\; i= 1,2,\cdots q, $
of the equations (\ref{beta1}-\ref{final1}) are momentum
dependent.

This procedure reminds Nested
Bethe Ansatz where we have a nested chain of Bethe equations (see as an
example the set of nested Bethe equations in t-J-\cite{EK} and staggered t-J
models \cite{TS}).

Although we have increased the amount of Bethe equations
(\ref{beta1})-(\ref{final1}) together with the
amount of parameters to be defined, each of the equations
(\ref{beta1}- \ref{final1}) now is
much simpler and allows to apply the technique of Thermodynamic Bethe
Ansatz \cite{Takahashi, Gauden} for their investigation.
After taking logarithm of the left and right hand sides
of these equations, one can reduce
them to the set of integral equations in the most interesting limit
of $q\rightarrow \infty$,
necessary in the irrational flux case. This fact justifies the procedure
of nested Bethe Ansatz.

Nested Bethe equations (\ref{beta1}- \ref{final1}) now have a form, which allows
to investigate them further by use of Thermodynamic Bethe Ansatz. One can take
logarithm of the left and right hand sides and transform it to integral
equation in the thermodynamic limit $q\rightarrow\infty$.

\section*{Acknowledgments} A.S acknowledges ISSP at the University of Tokyo
for hospitality where this work was initiated and carried out.

\appendix
\section{Nesting of Bethe Ansatz Equations}
\indent

Since any term in the linear space of functions $\Theta^{(q)} \in \Th_q$ can be
represented as a product of theta functions (\ref{basis}) with some particular
choice of zeros at $z_j, j=1,2, \cdots q$, the sum of the first and second
products of theta functions in the expression (\ref{eq:B}) can be represented
as
\begin{eqnarray}
\label{beta}
(-1)^p\beta_+ e^{i\frac{\Phi}{6}-\frac{2\pi i k p}{q}}
    \prod_{i=1}^q \Theta_1\Big(z-z_i+{p\over  q},\tau\Big)+
\qquad \quad \quad \no\\
 +(-1)^p\beta_-e^{-i\frac{\Phi}{6}+\frac{2\pi i k p}{q}}\prod_{i=1}^q
    \Theta_1\Big(z-z_i-{p\over  q},\tau\Big) \quad \quad \no\\
 = \gamma_{1} \prod_{i=1}^q \Theta_1\Big(z-z^{(1)}_i,\tau\Big).\quad \;\;
\end{eqnarray}
 Consequently, the sum of the third and forth terms in (\ref{eq:B})
(respectively the sum of fifth and sixth terms) define
\begin{eqnarray}
\label{alpha}
\alpha_+ e^{-i\frac{\Phi}{6}+\pi i\tau\frac{-q+1+2k}q-2\pi i z}
        \prod_{i=1}^q \Theta_1\Big(z-z_i-{\tau\over  q},\tau\Big)+
\quad  \no\\
+\alpha_-e^{i\frac{\Phi}{6}+ \pi i\tau\frac{q+1-2k}q+2\pi i z}
        \prod_{i=1}^q \Theta_1\Big(z-z_i+{\tau\over  q},\tau\Big) \quad \;\;\no\\
 = \gamma_{2} \prod_{i=1}^q \Theta_1\Big(z-z^{(2)}_i,\tau\Big),\quad
\end{eqnarray}

\begin{eqnarray}
\label{alphabeta} (-1)^p\delta_+e^{i\frac{\Phi}{3}+\pi i\tau\frac{1-2k+q}{q}+2\pi i z
-\frac{2\pi i k p}{q}}
\qquad \qquad \no\\
 \cdot\prod_{i=1}^q \Theta_1\Big(z-z_i+{\tau+p\over  q},\tau\Big)\;\;\; \no\\
+ (-1)^p \delta_-e^{i\frac{2\Phi}{3}+\pi i\tau\frac{1+2k-q}{q}-2\pi i z
+\frac{2\pi i k p}{q}}
\qquad \qquad \no\\
 \cdot\prod_{i=1}^q \Theta_1\Big(z-z_i-{\tau+p\over  q},\tau\Big)\;\; \\
 = \gamma_{3} \prod_{i=1}^q \Theta_1\Big(z-z^{(3)}_i,\tau\Big),\;\; \no
\end{eqnarray}

Right hand sides of the equations (\ref{beta}), (\ref{alpha}) and (\ref{alphabeta})
become zero at $z=z^{(1)}_j,\;z=z^{(2)}_j, \;z=z^{(3)}_j, \;j=1,2 \cdots q$,
respectively, which imposes restrictions on the parameters $z_i,  z^{(1)}_i,
z^{(2)}_i, z^{(3)}_i, i= 1,2,\cdots q$ in the form of Bethe equations
(\ref{beta1} - \ref{alphabeta1}).

The parameters $\gamma_1, \gamma_2$ and $\gamma_3$ are $z$ independent
constants, as follows from the arguments presented in the previous section by
use of representation of elliptic functions via Weierstrass $\zeta$ function
(\ref{eq:decom}). Therefore they can be fixed by inserting any convenient value
$z$ into the corresponding equation. By putting values $z=z_1+p/q,\;
z=z_1-\tau/q,\;z=z_1+p/q+\tau/q $ into the equations (\ref{beta}), (\ref{alpha})
and (\ref{alphabeta}) respectively, which annulate one of terms in the left
hand sides, one can solve this equations for $\gamma$'s and obtain
\begin{eqnarray}
\label{gama1}
\gamma_1 &=& (-1)^p\beta_+ e^{i\Phi(\frac{1}{6}-k)}\no\\
&\cdot& \frac{\prod_{i=1}^q \Theta_1\big(z_1-z_i+{2p\over  q},\tau\big)}
    {\prod_{i=1}^q \Theta_1\big(z_1-z^{(1)}_i+{p\over  q},\tau\big)}  \\
\label{gama2}
  \gamma_2 &=& \alpha_+ e^{-i\frac{\Phi}{6}+\pi i\tau\frac{3-q+2k}q-2\pi i z_1} \no\\
&\cdot& \frac{\prod_{i=1}^q \Theta_1\big(z_1-z_i-{2 \tau\over  q},\tau\big)}
        {\prod_{i=1}^q \Theta_1\big(z_1-z^{(2)}_i-{\tau\over  q},\tau\big)} \\
\label{gama3}
  \gamma_3 &=&(-1)^p\delta_+e^{i\Phi(\frac{4}{3}-k)+\pi
i\tau\frac{3-2k+q}{q}+2\pi i z_1} \no\\
 &\cdot&\frac{\prod_{i=1}^q \Theta_1\big(z_1-z_i+{2(\tau+p)\over q},\tau\big)}
 {\prod_{i=1}^q \Theta_1\big(z_1-z^{(3)}_i+{(\tau+p)\over q},\tau\big)}
\end{eqnarray}

Now let us continue further this procedure, which reminds nesting Bethe Ansatz
in the some of integrable models \cite{EK,EK1,TS}, and transform the sum of right hand sides
of (\ref{beta}) and (\ref{alpha}) into the product of theta functions with the zeros
at the new positions $z^{(12)}_i$
\vspace{2cm}

\begin{eqnarray}
\label{12}
&&\gamma_{1} \prod_{i=1}^q \Theta_1(z-z^{(1)}_i,\tau)+
\gamma_{2} \prod_{i=1}^q \Theta_1(z-z^{(2)}_i,\tau)\no\\
&=&\bar\gamma \prod_{i=1}^q \Theta_1(z-z^{(12)}_i,\tau)
\end{eqnarray}
Now again, one can insert the value $z=z^{(2)}_1$ into the
equation above and obtain
\begin{eqnarray}
\label{gamafinal}
\bar\gamma=\gamma_1 \frac{\prod_{i=1}^q \Theta_1\big(z^{(2)}_1-z^{(1)}_i,\tau\big)}
{\prod_{i=1}^q \Theta_1\big(z^{(2)}_1-z^{(12)}_i,\tau\big)},
\end{eqnarray}
where $\gamma_1$ defined by the expression in (\ref{gama1}).
The spectral parameters $z^{(12)}_i$ are the zeros of the
left(and, therefore, of the right) hand side of the equation
(\ref{12}). Hence, by use of (\ref{gama1}) and (\ref{gamafinal})
we obtain Bethe equations (\ref{gamagama}) for $z^{(12)}$.

With this reformulations of the sums of theta functions we
bring Schr\"odinger equation (\ref{eq:B}) for the
eigenenergy ${\cal E}$ to the very simple
form
\begin{eqnarray}
\label{final}
({\cal E}-3)\prod_{i=1}^q \Theta_1(z-z_i,\tau)=
\bar\gamma \prod_{i=1}^q \Theta_1\big(z-z^{(12)}_i,\tau\big)\no\\
+\gamma_{3} \prod_{i=1}^q \Theta_1\big(z-z^{(3)}_i,\tau\big)
\end{eqnarray}

By inserting $z=z^{(12)}_1$ into this expression
(the first term on the right hand side becomes zero) and using
(\ref{gama3}) for $\gamma_3$ one can obtain simple expression
(\ref{final0}) for the energy. This complets Bethe Ansatz solution of the
Hofstadter problem on the honeycomb lattice.

The set of nested Bethe equations (\ref{beta1} - \ref{final1}), are equivalent
to the single set (\ref{BAE1}). We have more equations in (\ref{beta1} - \ref{final1})
than in (\ref{BAE1}), but their advantage is in their simplicity. Each of mentioned
equations contains only two product of theta functions, which will allow
to implement the technique of Thermodynamic Bethe Ansatz \cite{Takahashi, Gauden}
for further investigations.

\end{document}